\documentclass{article}
\usepackage{spconf,amsmath,graphicx}

\usepackage{amsmath}
\usepackage{amssymb}
\usepackage{bbm}
\usepackage{physics}
\usepackage{siunitx}
\usepackage{trfsigns}
\usepackage{pifont}
\usepackage{etoolbox}
\usepackage{multirow}
\usepackage{csquotes}
\usepackage[shortlabels]{enumitem}  \usepackage{cite}   \usepackage{hyperref}
\usepackage[capitalize]{cleveref}
\usepackage{xifthen}
\usepackage{balance} \usepackage{booktabs}
\usepackage{listings}

\usepackage{url}

\let\oldcite\cite
\renewcommand{\cite}[1]{\ifthenelse{\isempty{#1}}{\inred{[cite!]}}{\oldcite{#1}}}

\newcolumntype{H}{>{\setbox0=\hbox\bgroup}c<{\egroup}@{}}      

  \newcommand{\tab}[2][c]{\begin{tabular}{@{}#1@{}}#2\end{tabular}}

\newcommand*{\thl}{\fontseries{b}\selectfont}
\robustify\thl
\robustify\fontseries

\newcommand{\inred}[1]{\textcolor{red}{#1}}

\usepackage[skip=7pt]{caption} 
 \usepackage{tikz}
\usepackage{pgfplots}
\pgfplotsset{compat=1.16}
\usepgfplotslibrary{groupplots}
\usetikzlibrary{positioning,arrows,matrix,fit,calc,patterns,chains,scopes,shapes.multipart,decorations,decorations.markings,backgrounds}

\usetikzlibrary {arrows.meta}
\usetikzlibrary{shapes.arrows}
\usetikzlibrary{intersections}

\definecolor{palette-1}{cmyk}{0.57, 0.27, 0, 0.16}    
\definecolor{palette-2}{HTML}{B1400D}    
\definecolor{palette-3}{cmyk}{0.35, 0, 0.59, 0.49}    
\definecolor{palette-4}{HTML}{8C0800}    
\definecolor{palette-5}{HTML}{591E71}    
\definecolor{palette-6}{HTML}{592F0D}    
\definecolor{palette-7}{HTML}{A23582}    
\definecolor{palette-8}{HTML}{3C3C3C}    
\definecolor{palette-9}{HTML}{B8850A}    
\definecolor{palette-10}{HTML}{006374}      

\tikzset{
line/.style={draw,black,thick,rounded corners=1mm,line cap=round},
    noshortarrow/.style={line,->},
    arrow/.style={noshortarrow,shorten >=.3mm},
    doublearrow/.style={arrow,<->, shorten <=.3mm},
box/.style={draw,black,thick,minimum height=3em,rounded corners=3,fill=white},
    nopadding/.style={minimum height=0,inner sep=1mm},
signalbox/.style={draw,black,thin,rounded corners=1mm,minimum width=7mm, minimum height=4mm,inner sep=0},
pbox/.style={box,fill=black!10},
backgroundbox/.style={inner xsep=3mm, inner ysep=1mm, draw, dashed, rounded corners,fill=orange!10},
branch/.style={inner sep=0.3mm,circle,fill=black},
operator/.style={draw,circle,black,rounded corners,inner sep=0,fill=white},
    vertex/.style={draw,ultra thin,circle,black,rounded corners,inner sep=0.6mm,fill=gray,fill opacity=0.5},
    edge/.style={line,very thick,line cap=butt},
pattern1/.style={pattern=north west lines,pattern color=palette-1},
    pattern2/.style={pattern=north east lines,pattern color=palette-2},
    pattern3/.style={pattern=crosshatch,pattern color=palette-3},
buswidth/.style={path picture={\draw[black,-] (path picture bounding box.south west) -- (path picture bounding box.north east);}}
}

\newcommand\defm[2]{\expandafter\newcommand{#1}{\ensuremath{{\textcolor{purple}{#2}}}}}

\defm\Loss{\mathcal{L}}

 \usepackage[acronym,shortcuts]{glossaries}
\glsdisablehyper

\newacronym{ASR}{ASR}{Automatic Speech Recognition}
\newacronym{CSS}{CSS}{Continuous Speech Separation}
\newacronym{cpWER}{cpWER}{Concatenated Minimum Permutation WER}
\newacronym{DER}{DER}{Diarization Error Rate}
\newacronym{LSTM}{LSTM}{Long Short Term Memory}

\newacronym{MTRNNT}{MT-RNN-T}{Multi-turn Recurrent Neural Network Transducer}
\newacronym{TSOT}{T-SOT}{Token-Level Serialized Output Training}
\newacronym{ORCWER}{ORC WER}{Optimal Reference Combination WER}

\newacronym{VAD}{VAD}{Voice Activity Detection}
\newacronym{WER}{WER}{Word Error Rate}

\newacronym{RIR}{RIR}{Room Impulse Response}

\title{Meeting Recognition with Continuous Speech Separation and Transcription-Supported Diarization}
\newcommand{\upb}{$^1$}
\newcommand{\ntt}{$^2$}
\name{\tab{Thilo von Neumann\upb, Christoph Boeddeker\upb, Tobias Cord-Landwehr\upb, Marc Delcroix\ntt,\\Reinhold Haeb-Umbach\upb}}
\address{\upb Paderborn University\quad \ntt NTT corporation, Japan}
\begin{document}
\ninept
\maketitle
\begin{abstract}
We propose a modular pipeline for the single-channel separation, recognition, and diarization of meeting-style recordings and evaluate it on the Libri-CSS dataset.
Using a Continuous Speech Separation (CSS) system with a TF-GridNet separation architecture, followed by a speaker-agnostic speech recognizer, we achieve state-of-the-art recognition performance in terms of Optimal Reference Combination Word Error Rate (ORC WER).
Then, a d-vector-based diarization module is employed to extract speaker embeddings from the enhanced signals and to assign the CSS outputs to the correct speaker.
Here, we propose a syntactically informed diarization using sentence- and word-level boundaries of the ASR module to support speaker turn detection.
This results in a state-of-the-art Concatenated minimum-Permutation Word Error Rate (cpWER) for the full meeting recognition pipeline.

\end{abstract}
\section{Introduction}
\label{sec:intro}

Automatic meeting transcription aims at answering the question ``who spoke what and when'' \cite{Raj2021}.
To do so, the tasks of speech separation, diarization and recognition have to be solved. 
Integrated and modular architectures have been proposed for meeting transcription.

Integrated solutions, such as the \gls{MTRNNT} \cite{Sklyar2022_MultiTurnRNNTStreaming} and the \gls{TSOT} \cite{Kanda2022_Streaming} use a transformer transducer model to recognize the speech of all speakers in a multi-talker recording without an explicit separation module. Speaker assignment is then performed on the segmented recognition output. 
The attractiveness of those integrated approaches is the training under a common objective function, which potentially delivers superior results.

In contrast, modular systems combine modules that explicitly perform speech separation, diarization, and recognition.
They have the advantage that the individual modules can be optimized independently and implemented using different tools.
However, it is unclear which processing order of the aforementioned three tasks is the most appropriate.

In \cite{Raj2021}, three variants of the modular approach were introduced. 
The \enquote{CHiME-6 pipeline} first conducts diarization, followed by separation and then \gls{ASR} \cite{Watanabe2020_CHiME6ChallengeTackling}.
This pipeline benefits from the wealth of literature on diarization, where in recent years, effective solutions even for partially overlapping speech have been found \cite{Medennikov2020, 21_bredin_segmentation}.
The diarization output can guide the source separation \cite{Boeddeker_2018}, or the speaker information can be used to perform target speaker extraction, e.g., using SpeakerBeam \cite{delcroix2021speaker,Zmolikova_2019}.
However, poor diarization can negatively impact the source separation and, thus, recognition accuracy.

Therefore, Raj et al. proposed to swap the order of diarization and separation to conduct separation before diarization \cite{Raj2021}. 
This framework employs \gls{CSS}, which is a source separation scheme that maps multi-talker speech with an arbitrary number of speakers to a fixed number of output channels (typically two) such that there is no speaker overlap on any of the output channels \cite{Chen2020_ContinuousSpeechSeparation}. 
We refer thus to this pipeline as CSS followed by diarization and \gls{ASR} (CSS-DA).
By performing separation first, we can greatly simplify the diarization system since it does not need to handle overlap. 
For example, (conventional) clustering-based diarization can be sufficient if all speech overlaps are removed.
The diarization performance depends, however, on the quality of the source separation. 
Furthermore, segmentation remains an issue as it is not always possible to reliably detect speaker changes just from the speech activity, especially when speakers take turns quickly. 

The third processing order first carries out speech recognition on the CSS output followed by diarization on the \gls{ASR} output, which we call CSS-AD to emphasize the processing order.\footnote{It was called ``CSS pipeline'' in \cite{Raj2021}.}  
It enjoys the same advantages as the CSS-DA pipeline. In addition, the diarization may benefit from word boundary information to, e.g., reduce false alarms if a robust recognizer is employed.

Although the third option appears appealing since diarization does not need to care for overlap and can exploit information from the ASR system, only a few studies have followed this pipeline\cite{Yoshioka2019}.
In \cite{dimitriadis17_interspeech}, timestamps from ASR were used for diarization.
However, they did not integrate it into a full CSS pipeline since they did not use separation and thus could not handle overlapped speech. 
The full CSS-AD pipeline has been implemented in \cite{Yoshioka2019}, but it heavily relied on multi-channel and multi-modal inputs to cope with imperfect separation or artifacts introduced by the separation stage.

\begin{figure*}[tb]
    \centering
    \begin{tikzpicture}[x=1.5em]

\pgfdeclarelayer{bg-1}
\pgfdeclarelayer{bg-0}
\pgfdeclarelayer{fg-1}
\pgfdeclarelayer{fg-0}
\pgfsetlayers{background,bg-1,fg-1,bg-0,fg-0,main}

\tikzset{>=stealth}

    \foreach \i in {0} {
        \coordinate (blockshift) at (\i*0.2,\i*0.2);
            
    \begin{pgfonlayer}{fg-\i}
        \node[box] (separator) {TF-GridNet};
        \node[box,left=1 of separator, align=center] (segment) {Uniform \\Segmentation};
        \node[box,right=1 of separator] (sticher) {Stitching};

        \node[box, anchor=west, align=center] (vad) at ($(sticher.east)+(1,0) + (0,0) +(blockshift)$) {VAD-based\\Segmentation};
        \node[box,right=1 of vad] (asr) {ASR};
        \node[box,right=1 of asr] (subseg) {Sub-segmentation};
        \node[box,right=1 of subseg] (dvec) {d-Vector};
        \node[box,right=1 of dvec] (clustering) {k-Means};
    
        \draw[arrow] (segment) -- (separator);
        \draw[arrow] (separator) -- (sticher);
        \draw[arrow] (sticher.east|-vad) -- (vad);
        \draw[arrow] (vad) -- (asr);
        \draw[arrow] (asr) -- (subseg);
        \draw[arrow] (subseg) -- (dvec);
        \draw[arrow] (dvec) -- (clustering);

        \draw[arrow] (segment.west) + (-1, 0) -- ++(0, 0);
        \draw[arrow] (clustering.east) -- ++(1, 0);

        \node[above=0.1em] (csslabel) at ($(segment.north west)!1/2!(sticher.north east)$) {Continuous Speech Separation (CSS)};
        
        \node[above=0.1em] (dialabel) at ($(subseg.north west)!1/2!(clustering.north east)$) {\vphantom{Ag} Diarization};
    \end{pgfonlayer}

    \def\boxsep{0.55em}
    \begin{pgfonlayer}{bg-1}
        
        \node[fit={(separator)(segment)(sticher)(csslabel)}, inner sep=\boxsep,fill=gray!10,draw=gray,rounded corners] (css) {};

        \coordinate (tmp) at ($(asr.west)-(asr.west)!0.5!(vad.east)$);
    \end{pgfonlayer}
    
    \begin{pgfonlayer}{bg-\i}

            \draw[fill=gray!10,draw=gray,rounded corners]
                let \p{tmp}=(tmp) in
                ($(subseg.south west) + (-\x{tmp},-\boxsep) + (blockshift)$)
                -|
($(dialabel.north-|clustering.east) + (\boxsep,\boxsep) + (blockshift)$)
                -|
                ($(vad.south west) + (-\boxsep,-\boxsep) + (blockshift)$)
                -|
                ($(vad.north east) + (\x{tmp},\x{tmp}) + (blockshift)$)
                -|
                cycle
                ;

    \end{pgfonlayer}

    }

\end{tikzpicture} 

    \caption{Proposed processing pipeline for meeting transcription.}
    \label{fig:pipeline}
\end{figure*}
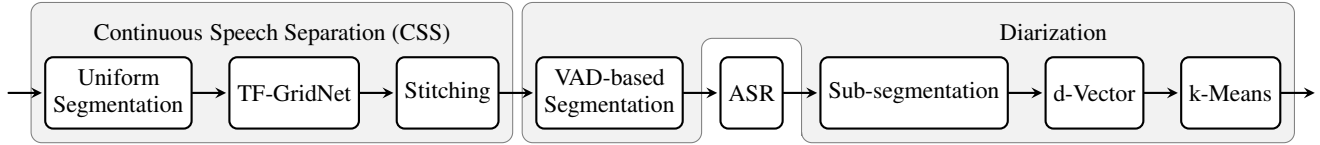

In this paper, we investigate the potential of the CSS-AD pipeline for processing single-channel meeting recordings.
First, we implement a single-channel CSS using TF-GridNet\cite{Wang2022_TFGridNetIntegratingFull, Wang2022_TFGridNetMakingTimeFrequency}. Although TF-GridNet has been shown to achieve a state-of-the-art separation performance for fully overlapped speech on reverberant mixtures\cite{Wang2022_TFGridNetIntegratingFull, Wang2022_TFGridNetMakingTimeFrequency}, it has not been investigated with partially overlapped speech and CSS-style separation of long recordings. 
We demonstrate that it indeed has the potential for CSS, but diarization after separation remains challenging because speech activity is not enough to detect speaker turns. 
We introduce a novel segmentation approach that exploits syntactic information, such as word and sentence boundaries, from the ASR system. 
In particular, since speaker embeddings benefit from long speech segments, we propose to segment the separated signals at sentence boundaries predicted by the recognizer. 
Due to imperfect separation, sentences can contain speech from multiple speakers.
We thus refine these segments by detecting speaker changes by word-level speaker embeddings.

We carry out experiments on the well-known Libri-CSS meeting recognition task \cite{Chen2020_ContinuousSpeechSeparation}. 
We confirm that the TF-GridNet architecture is well-suited for CSS and achieves state-of-the-art \gls{ORCWER} on the task. 
In addition, using our proposed transcription-supported diarization, we also achieve state-of-the-art \gls{cpWER}, surpassing prior work by a relative WER improvement of 20 \%. These strong results suggest the potential for the modular CSS-AD pipeline, which performs diarization after separation and ASR, and may encourage more research in that direction.

\section{Continuous Speech Separation}
A block diagram of the proposed CSS-AD processing pipeline is shown in \Cref{fig:pipeline}. It consists of \gls{CSS}, followed by \gls{VAD}-based segmentation, \gls{ASR}, and finally clustering-based diarization.
\subsection{TF-GridNet}
We use TF-GridNet \cite{Wang2022_TFGridNetIntegratingFull,Wang2022_TFGridNetMakingTimeFrequency} to achieve state-of-the-art separation performance. 
The TF-GridNet works in the STFT domain and uses spectral mapping instead of masking.
It alternates between modeling cross-time dependencies for every frequency, cross-frequency dependencies for each time frame, and global dependencies.
Cross-time and cross-frequency dependencies are modeled with \glspl{LSTM} and global dependencies are modeled with attention.

TF-GridNet has been applied to fully overlapped speech separation and speech enhancement \cite{Wang2022_TFGridNetIntegratingFull}, but not to conversational speech, where it is important to separate speech in overlapping regions and not negatively affect single-speaker regions.

To this end, we trained the separator on fully overlapped two-speaker mixtures and single-speaker data that we generated with the MMS-MSG tool \cite{Cord2022_MMSMSG}.
We use the configuration suggested in \cite{Wang2022_TFGridNetIntegratingFull} for reverberant speech separation, i.e., $B=4$ blocks and a hidden size of $H=192$.

\subsection{Continuous processing with stitching}
\glsreset{CSS}

Extending TF-GridNet from fully overlapping utterances to long conversations that exceed minutes and contain many speakers can be done with the \gls{CSS} concept proposed in \cite{Yoshioka2018_CSS}, which is also frequently referred to as stitching.

The two-speaker separation system (here TF-GridNet) is applied independently to short,  overlapping segments of the recording, here \SI{4}{\second} long with \SI{2}{\second} overlap.
This results in an inter-segment speaker-permutation problem. Neighboring segments thus have to be aligned after separation to obtain continuous and consistent output signals from concatenation. The permutation is found based on an MSE criterion computed on the overlapping parts of two adjacent segments.

With this approach, a two-speaker separator can be applied to an arbitrarily long recording containing arbitrarily many speakers.
However, given only two output channels, it is constrained to at most two speakers active within a single segment.
The outputs of this model are two overlap-free streams, with each stream containing the speech of multiple speakers.
Therefore, diarization is not yet solved at this stage.

\section{Diarization and ASR}
For diarization, we apply a  modular clustering-based diarization pipeline composed of \gls{VAD}, embedding estimation, and k-Means clustering.
This pipeline is then extended with a sub-segmentation module that allows the detection of speaker turns in the \gls{VAD} segments in order to attribute these regions to the correct speaker.
The sub-segmentation is supported by timing information given by the \gls{ASR} engine.
 \cref{fig:segmentation} illustrates these sub-segmentation approaches.

\subsection{VAD-based Segmentation}
\glsreset{VAD}
An energy-based \gls{VAD}, as shown in \cref{fig:segmentation}-(a), is employed to detect utterance boundaries in the separated signals.
The framewise energy is computed in the time-frequency domain and thresholding is applied to detect regions of active speech.
Then, the estimate is smoothed over time using morphological closing (dilation followed by erosion).
The parameters of the closing operation are chosen such that speech activity is over-estimated by \SI{0.4}{\second}, which improves \gls{ASR} performance \cite{Boeddeker2023_TSSEP}.

\subsection{Automatic Speech Recognition}
\glsreset{ASR}
We employed Whisper \cite{Radford2023_Whisper} for \gls{ASR} because its output is accompanied by word boundary timestamps. 
This information is used to support the segmentation, which is an essential component of the diarization. 

\subsection{Sub-segmentation}
\label{sec:segmentation}
\begin{figure}[tb]
    \centering

\begin{tikzpicture}[y=2.5em]

\newcommand{\backgroundColor}{white}

\tikzset{
	word/.style={
execute at begin node={\vphantom{Ag}},
		execute at end node={\vphantom{Ag}},
		inner sep=0.16em,
		font = {\scriptsize},
		line width=0,
		anchor=west,
	},
	spkOne/.style={palette-1},
	spkTwo/.style={palette-2},
	spkThree/.style={palette-3},
spkDiaOne/.style={draw=none,fill=palette-1},
spkDiaTwo/.style={draw=palette-2,pattern=crosshatch,pattern color=palette-2,draw=none,preaction={clip,postaction={draw=palette-2, line width=0.15em}}},
spkDiaThree/.style={draw=palette-3!60,pattern=north west lines,pattern color=palette-3!60,draw=none,preaction={clip,postaction={draw=palette-3!60, line width=0.15em}}},
	outputScope/.style={x=1em, y=-.7em},
	outputBoundingBox/.style={draw=black, fill=gray!40},
	segmentationlabel/.style={above=0.5em,font={\scriptsize},inner sep=0,fill=\backgroundColor},
	branch/.style={inner sep=0.07em,circle,fill=black},
     segmentarrow/.style={{|<[scale=0.5]}-{>[scale=0.5]|[scale=0.5]}},
}
\tikzset{>=Stealth}

	\node[word,spkOne] (word-1) at (0, 0) {Hello,};
	\node[word,spkOne] (word-2) at ($(word-1.east)+(0,0)$) {let’s};
	\node[word,spkOne] (word-3) at ($(word-2.east)+(0,0)$) {start};
\node[word,spkOne] (word-4) at ($(word-3.east)+(0,0)$) {uh};
	\node[word,spkTwo] (word-5) at ($(word-4.east)+(0,0)$) {sorry};
	\node[word,spkTwo] (word-6) at ($(word-5.east)+(0,0)$) {I’m};
	\node[word,spkTwo] (word-7) at ($(word-6.east)+(0,0)$) {late};
	\node[word,spkOne] (word-8) at ($(word-7.east)+(0,0)$) {uh};
	\node[word,spkOne] (word-9) at ($(word-8.east)+(0,0)$) {the};
	\node[word,spkOne] (word-10) at ($(word-9.east)+(0,0)$) {meeting.};
	\node[word,spkThree] (word-11) at ($(word-10.east)+(0,0)$) {What};
	\node[word,spkThree] (word-12) at ($(word-11.east)+(0,0)$) {do};
	\node[word,spkThree] (word-13) at ($(word-12.east)+(0,0)$) {we};
	\node[word,spkThree] (word-14) at ($(word-13.east)+(0,0)$) {discuss};
	\node[word,spkThree] (word-15) at ($(word-14.east)+(0,0)$) {today?};

	\begin{scope}[shift={(0,-0.7)}]

		\draw[spkDiaOne] (word-1.west|-0,1ex) rectangle (word-4.east|-0,-0ex);
		\draw[spkDiaTwo] (word-5.west|-0,1ex) rectangle (word-7.east|-0,0ex);
		\draw[spkDiaOne] (word-8.west|-0,1ex) rectangle (word-10.east|-0,0ex);
		\draw[spkDiaThree] (word-11.west|-0,1ex) rectangle (word-15.east|-0,0ex);
		
		\node[above right,font={\tiny}] () at (0,0.3ex) {\textit{Reference diarization}};
	\end{scope}
	
	\newcommand{\blockshift}{-1}

	\begin{scope}[shift={($(0,-1.5em)+(0,\blockshift)$)},y=1.5em]
		
		\node[segmentationlabel] (vadlabel) at ($(word-1.west|-0,0)!0.5!(word-15.east|-0,0)$) {(a) VAD segmentation};
	
		\draw[segmentarrow] (word-1.west|-0,0) -- (word-15.east|-0,0);
		\begin{scope}[shift={(0,-1)}]
			\draw[spkDiaOne] (word-1.west|-0,0.5ex) rectangle (word-15.east|-0,-0.5ex);
			\node[above right,font={\tiny}] () at (0,0) {\textit{Diarization}};
		\end{scope}

	\draw[->,-{Triangle[width=1.5ex,length=1ex]},line width=0.8ex,gray] ($(word-1.west|-0,-0.2em)!5/9!(word-15.west|-0,-0.2em)$) -- +($(0,-1)+(0,0.5em)$);
		
		\coordinate (vadBottom) at ($(0,-1)+(0,-0.5ex)$);
	\end{scope}
	
	\begin{scope}[shift={($(vadBottom)+(0,\blockshift)$)},y=1.5em]
	
		\node[segmentationlabel] (unilabel) at ($(word-1.west|-0,0)!0.5!(word-15.east|-0,0)$) {(b) Uniform sub-segmentation};
	
		\foreach \start/\stop/\color in {
			0/0.2/spkDiaOne,
			0.2/0.4/spkDiaTwo,
			0.4/0.6/spkDiaOne,
			0.6/0.8/spkDiaThree,
			0.8/1/spkDiaThree}{
			\draw[segmentarrow] ($(word-1.west|-0,0)!\start!(word-15.east|-0,0)$) -- ($(word-1.west|-0,0)!\stop!(word-15.east|-0,0)$);
			\begin{scope}[shift={(0,-1)}]
				\draw[\color] ($(word-1.west|-0,0.5ex)!\start!(word-15.east|-0,0.5ex)$) rectangle ($(word-1.west|-0,-0.5ex)!\stop!(word-15.east|-0,-0.5ex)$);
			\end{scope}
			
		}
		\node[above right,font={\tiny},black] () at (0,-1) {\textit{Diarization}};

		\draw[->,-{Triangle[width=1.5ex,length=1ex]},line width=0.8ex,gray] ($(word-1.west|-0,-0.2em)!5/9!(word-15.west|-0,-0.2em)$) -- +($(0,-1)+(0,0.5em)$);
		
		\coordinate (uniBottom) at ($(0,-1)+(0,-0.5ex)$);
	\end{scope}
	
	\begin{scope}[shift={($(uniBottom)+(0,\blockshift)$)},y=1.5em]
	
		\node[segmentationlabel] (sentencelabel) at ($(word-1.west|-0,0)!0.5!(word-15.east|-0,0)$) {(c) Sentence-level sub-segmentation};
		
		\draw[segmentarrow] (word-1.west|-0,0) -- (word-10.east|-0,0);
		\draw[segmentarrow] (word-11.west|-0,0) -- (word-15.east|-0,0);
		\begin{scope}[shift={(0,-1)}]
			\draw[line width=1ex,spkDiaOne] (word-1.west|-0,0.5ex) rectangle (word-10.east|-0,-0.5ex);
			\draw[line width=1ex,spkDiaThree] (word-11.west|-0,0.5ex) rectangle (word-15.east|-0,-0.5ex);
			\node[above right,font={\tiny}] () at (0,0) {\textit{Diarization}};
		\end{scope}
	
		\draw[-{>[scale=0.5]}] (word-11.west|-0,0) -- node[right,inner sep=0.1em,font={\tiny}](){SB} +($(0, -1)+(0,0.5ex)$);
		
		\draw[->,-{Triangle[width=1.5ex,length=1ex]},line width=0.8ex,gray] ($(word-1.west|-0,-0.2em)!5/9!(word-15.west|-0,-0.2em)$) -- +($(0,-1)+(0,0.5em)$);
	
		\coordinate (sentenceBottom) at ($(0,-1)+(0,-0.5ex)$);
	\end{scope}
	
	\begin{scope}[shift={($(sentenceBottom)+(0,\blockshift)$)},y=1.5em]
	
		\node[segmentationlabel] (wordlabel) at ($(word-1.west|-0,0)!0.5!(word-15.east|-0,0)$) {(d) Word-level sub-segmentation};
	
		\foreach \i in {1,...,15}{
			\draw[segmentarrow] (word-\i.west|-0,0) -- (word-\i.east|-0,0);
		}
	
		\begin{scope}[shift={(0,-1.8)},y=1.5em]
			\draw[->] (word-1.west|-0,0) -- ($(word-15.east|-0,0)+(0,0)$);
			\draw[->] (word-1.west|-0,0) +(0,-0.1) -- +(0,1) node[right,font=\tiny,yshift=-0.5ex] () {Word-level cosine similarity};
		
			\foreach \i/\j in {1/0.5,2/0.6,3/0.55,4/0.15,5/0.45,6/0.75,7/0.1,8/0.3,9/0.6,10/0.12,11/0.6,12/0.7,13/0.65,14/0.65}{
				\node[branch] () at (word-\i.east|-0,\j) {};
			}
			\draw[dash pattern=on 1pt off 1pt] (word-1.west|-0,0.4) -- node[above=-0.2em,pos=0.60,font={\tiny}](){\textit{Threshold}} (word-15.east|-0,0.4);
		
			\begin{scope}[shift={(0,-1)}]
				\draw[segmentarrow] (word-1.west|-0,0) -- (word-4.east|-0,0);
				\draw[segmentarrow] (word-5.west|-0,0) -- (word-7.east|-0,0);
				\draw[segmentarrow] (word-8.west|-0,0) -- (word-10.east|-0,0);
				\draw[segmentarrow] (word-11.west|-0,0) -- (word-15.east|-0,0);
				
				\draw[-{>[scale=0.5]}] (word-11.west|-0,1) -- node[pos=0.75,right,inner sep=0.1em,font={\tiny}](){SC} +($(0, -2)+(0,0.5ex)$);
				\draw[-{>[scale=0.5]}] (word-8.west|-0,1) -- node[pos=0.75,right,inner sep=0.1em,font={\tiny}](){SC} +($(0, -2)+(0,0.5ex)$);
				\draw[-{>[scale=0.5]}] (word-5.west|-0,1) -- node[pos=0.75,right,inner sep=0.1em,font={\tiny}](){SC} +($(0, -2)+(0,0.5ex)$);
				
				\foreach \i in {1,...,14}{
					\draw[] (word-\i.east|-0,1.1) -- (word-\i.east|-0,0.9);
				}
				
				\begin{scope}[shift={(0,-1)}]
					\draw[spkDiaOne] (word-1.west|-0,0.5ex) rectangle (word-4.east|-0,-0.5ex);
					\draw[spkDiaTwo] (word-5.west|-0,0.5ex) rectangle (word-7.east|-0,-0.5ex);
					\draw[spkDiaOne] (word-8.west|-0,0.5ex) rectangle (word-10.east|-0,-0.5ex);
					\draw[spkDiaThree] (word-11.west|-0,0.5ex) rectangle (word-15.east|-0,-0.5ex);
					
					\node[above right,font={\tiny}] () at (0,0) {\textit{Diarization}};

					\coordinate (wordBottom) at (0,-0.5ex);
				\end{scope}
				\draw[->,-{Triangle[width=1.5ex,length=1ex]},line width=0.8ex,gray] ($(word-1.west|-0,-0.2em)!5/9!(word-15.west|-0,-0.2em)$) -- +($(0,-1)+(0,0.5em)$);
			\end{scope}
		\end{scope}

		\begin{pgfonlayer}{background}
		
			\def\boxsep{0.3em}
			
			\node[fit={(vadlabel)(vadBottom)(vadBottom-|word-1.west)(vadBottom-|word-15.east)}, inner sep=\boxsep,fill=\backgroundColor,draw=gray,rounded corners] () {};
			\node[fit={(unilabel)(uniBottom)(uniBottom-|word-1.west)(uniBottom-|word-15.east)}, inner sep=\boxsep,fill=\backgroundColor,draw=gray,rounded corners] () {};
			\node[fit={(sentencelabel)(sentenceBottom)(sentenceBottom-|word-1.west)(sentenceBottom-|word-15.east)}, inner sep=\boxsep,fill=\backgroundColor,draw=gray,rounded corners] () {};
			\node[fit={(wordlabel)(wordBottom)(wordBottom-|word-1.west)(wordBottom-|word-15.east)}, inner sep=\boxsep,fill=\backgroundColor,draw=gray,rounded corners] () {};
		
			\draw[dash pattern=on 1pt off 1pt, gray!50] (word-1.north west) -- (word-1.north west|-wordBottom);
			\draw[dash pattern=on 1pt off 1pt, gray!50] (word-5.north west) -- (word-5.north west|-wordBottom);
			\draw[dash pattern=on 1pt off 1pt, gray!50] (word-8.north west) -- (word-8.north west|-wordBottom);
			\draw[dash pattern=on 1pt off 1pt, gray!50] (word-11.north west) -- (word-11.north west|-wordBottom);
			\draw[dash pattern=on 1pt off 1pt, gray!50] (word-15.north east) -- (word-15.north east|-wordBottom);

		\end{pgfonlayer}
		
	\end{scope}
	
\end{tikzpicture}     
    \caption{Comparison of the different segmentation schemes. Here, colors represent speakers. SB and SC stand for sentence boundary and speaker change.}
    \label{fig:segmentation}
\end{figure}
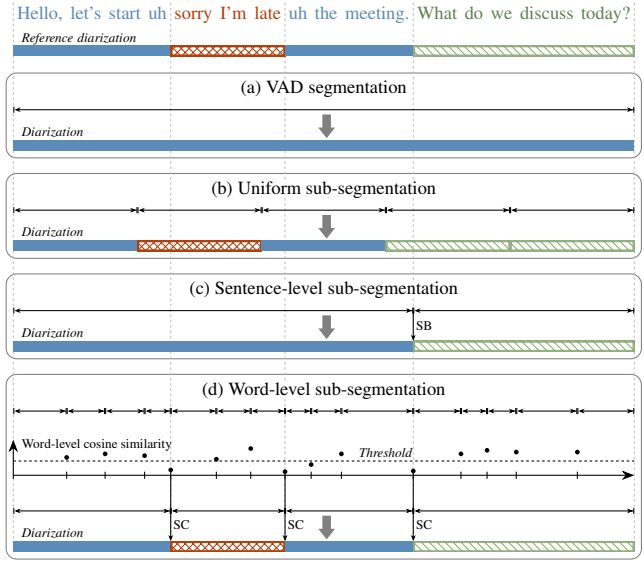

A \gls{VAD}-based segmentation is unable to distinguish between speakers and is, therefore, unable to detect speaker changes or quick speaker turns, resulting in activity segments that may contain more than a single speaker.
To mitigate this, we sub-segment each continuous speech region found by the initial \gls{VAD}. For the sub-segmentation, we explore different options.

\subsubsection{Uniform segmentation}
Uniform segmentation sub-segments the audio into equally sized pieces, as shown in \cref{fig:segmentation}-(b) and is commonly employed in diarization \cite{Garcia2017_Speaker}.
The choice of the segment length is a trade-off between temporal resolution (smaller segments give better resolution) and the speaker embedding quality (longer segments result in better embeddings while, here, also increasing the probability for mixed-speaker segments).
Typically, speaker embeddings are considered to become uninformative if computed on segments shorter than \SI{2}{\second} but stop to improve in quality if the length exceeds \SI{4}{\second} of audio \cite{Zhou2021_resnext}.
We here perform uniform segmentation after \gls{ASR} and move words that happen to lie on a sub-segment boundary to the sub-segment with which they most overlap.

\subsubsection{Sentence-level segmentation}
The uniform segmentation is not guaranteed to segment in positions that are beneficial for speaker clustering.
With the assumption that speaker changes are more likely at sentence boundaries than within a sentence, we propose to 
sub-segment at positions where the \gls{ASR} system predicts an end-of-sentence tag, e.g., \ a full stop. 
This improves the quality of the speaker embeddings extracted from the sub-segments because the detected segments are more likely to contain only a single speaker, while at the same time maximizing the segment length for more robust speaker embeddings.
Fig. \ref{fig:segmentation}-(c) shows an example of sentence-level sub-segmentation.

\subsubsection{Word-level d-vectors for segmentation}

Although speaker changes likely happen at the end of sentences, they may also happen in the middle of a sentence, or the \gls{ASR} model may fail to predict the end-of-sentence tag.
We implement the following procedure to find those cases.

The start and end times of words as predicted by the \gls{ASR} system are used to estimate word-level speaker embeddings, here d-vectors (see \cref{sec:dvector}), and the positions between words are treated as candidates for speaker-change-points.
For each speaker-change-point candidate, we compute a score as the cosine similarity between the average of the six preceding word-level embeddings and the six following word-level embeddings.
A candidate with a lower score is more likely to be a true speaker-change-point.
We use the average over a certain context because the d-vectors are noisy due to the usually short word durations.
We hypothesize a speaker change when a candidate's score is the lowest within a context of six words to the left and right and at the same time below a threshold of $0.2$.
Note that the word-level d-vectors are only used for sub-segment estimation and not for the follow-up clustering, which is described in the next section. 
Fig. \ref{fig:segmentation}-(d) shows an example of the word-level sub-segmentation.

When sentence-level segmentation is combined with word-level segmentation, the segments are first split using the sentence boundaries.
The resulting segments are then subdivided using the word-level d-vector approach.

\subsection{Speaker embedding Clustering}
\label{sec:dvector}

We use the speaker embedding extractor from  \cite{Cord-Landwehr2023_teacherstudent} to compute d-vectors.
This model is based on the ResNet-34 architecture \cite{Li2017_deep_speaker}, and is trained on VoxCeleb \cite{Nagrani2020_VoxcelebLargescaleSpeaker} segments, where noise from the MUSAN database \cite{Snyder2015_MUSAN} was added and signals were reverberated with simulated room impulse responses for data augmentation.
The d-vectors are extracted for each sub-segment and normalized to unit length. 
Then, they are clustered with k-Means++ \cite{07_arthur_kmeans} to obtain the speaker labels for each sub-segment. 
The diarization results consist of the combination of the sub-segment boundaries and the associated speaker labels.

\section{Experiments}
\label{sec:experiments}

\subsection{Data}
For the experiments, we used Libri-CSS \cite{Chen2020_ContinuousSpeechSeparation}, which is a commonly used database for the evaluation of meeting transcription systems.
It consists of \num{10} hours of meeting data that is simulated based on the LibriSpeech \verb|test_clean| set \cite{Panayotov2015_librispeech} with overlap ratios between \SIrange{0}{40}{\percent}.
These meetings were then re-recorded in a conference room with distributed loudspeakers in order to mimic
real meetings.

\subsection{ASR systems}
For the recognition, we used Whisper \cite{Radford2023_Whisper} in the \enquote{large-v2} configuration.
This recognizer ignores speaker changes and provides continuous transcriptions after turn-takings.
Additionally, it delivers punctuation symbols, from which the end of sentences can be detected, and the time stamps of the boundaries of each estimated word.
Both properties make it well suited for our proposed transcription-supported sub-segmentation.

While Whisper is a strong general \gls{ASR} model, it falls behind compared to in-domain trained LibriSpeech models: Whisper \enquote{large-v2} achieves 5.2\%  \gls{WER} on \verb|dev_other| while \cite{Chang2022end} reports 3.7\% on huggingface\footnote{\url{https://huggingface.co/espnet/simpleoier_librispeech_asr_train_asr_conformer7_wavlm_large_raw_en_bpe5000_sp}}. To push the pipeline's performance, we hence use the conformer-based ASR system from \mbox{ESPnet} that utilizes WavLM features \cite{Chen2022wavlm} as a second speech recognizer after diarization.
Note that this model does not deliver word boundaries and can thus not be used for transcription-supported diarization.

We use the off-the-shelf ASR systems without fine-tuning them to the artifacts produced by the separation.

\subsection{Metrics}
\glsreset{ORCWER}
\glsreset{cpWER}

We evaluate our meeting transcription pipeline using \gls{ORCWER} \cite{Sklyar2022_MultiTurnRNNTStreaming}, \gls{cpWER} \cite{Watanabe2020_CHiME6ChallengeTackling} using the MeetEval\footnote{\url{https://github.com/fgnt/meeteval}} toolkit,  and \gls{DER} \cite{Sadjadi2021_nist_sre_plan} computed with the \verb|md-eval-22.pl| script without a collar.

The \gls{ORCWER} assigns each reference utterance to one system output stream such that the \gls{WER} is minimized. 
It ignores speaker label errors.
In contrast, the \gls{cpWER} takes the assigned speaker labels into account and evaluates the WER for each speaker. 
Thus, it is the combination of both metrics that describes the \gls{ASR} performance and the overall capability of the pipeline.

\subsection{Evaluation without diarization}
First, we skip the diarization part and evaluate the pipeline without sub-segmentation and clustering in \cref{tab:orc}.
Without the speech separation front-end  and using only \gls{VAD}, the Whisper \gls{ASR} yields an \gls{ORCWER} of \SI{26.5}{\percent}.
By performing separation with TF-GridNet in the \gls{CSS} pipeline, we obtained an \gls{ORCWER} of \SI{6.8}{\percent}, which already beats any \gls{ORCWER} we found in the literature.
With ESPnet and the WavLM features, we got a further improvement of \SI{0.4}{\percent}, while Whisper achieved \SI{3.5}{\percent} when applied to the clean sources.

\Cref{tab:orc} also contains results from literature. While \cite{Sklyar2022_MultiTurnRNNTStreaming} reported an \gls{ORCWER} achieved with \gls{MTRNNT}, the other two cited numbers cannot directly be compared because they did not compute \gls{ORCWER}, but similar quantities, from which it can only be concluded that the \gls{ORCWER} is in the same ballpark.

\begin{table}
    \centering
    \caption{ORC WER on Libri-CSS. $^+$: cpWER with oracle speaker label assignment. *: asclite WER.}
    \label{tab:orc}
    \renewcommand*{\arraystretch}{1.1}
    \begin{tabular}{llS}
        \toprule
         Model & ASR & {ORC WER} \\
         \midrule
         No separation & Whisper & 26.5 \\
         TF-GridNet & Whisper & 6.8\\ 
         TF-GridNet & ESPnet & 6.4 \\ 
Clean signals & Whisper & 3.5 \\
         \midrule
         \gls{MTRNNT} \cite{Sklyar2022_MultiTurnRNNTStreaming} & Integrated & 23.6 \\
         t-SOT TT \cite{Kanda2022_Streaming} & Integrated & 7.6* \\
         TS-SEP \cite{Boeddeker2023_TSSEP} & ESPnet & 5.8$^+$ \\
         \bottomrule
    \end{tabular}
\end{table}

\subsection{Diarization Evaluation}
\Cref{tab:main} reports \glspl{cpWER} for the different sub-segmentation schemes discussed in Section \ref{sec:segmentation}. Using the segments found by the \gls{VAD} and performing clustering on the d-vectors computed from those segments gave a rather poor performance of \SI{14.8}{\percent} \gls{cpWER}.
We evaluated the oracle clustering performance to determine whether this was caused by bad segmentation or bad clustering. The oracle clustering attributes the speaker label that minimizes the \gls{cpWER}, as done in  \cite{Boeddeker2023_TSSEP}.
Oracle clustering only slightly improved cpWER to \SI{13.9}{\percent}, which indicates that the \gls{VAD} segmentation is the major problem.
With  uniform segmentation with \SI{4}{s} and with \SI{2}{s}  we achieved \SI{9.8}{\percent} and \SI{12.4}{\percent} \gls{cpWER}, respectively.
The comparison with the results achieved with oracle clustering shows that shorter segments are, in principle, better if only reliable d-vectors could be obtained. Our proposed sub-segmentation schemes based on syntactic information significantly improved the \gls{cpWER}, and using first sentence-level information, and then the word-level d-vectors pushed the performance down to \SI{7.2}{\percent} \gls{cpWER}.

Table \ref{tab:der_literature} compares the DER and cpWER of our proposed method with those of prior works.
We selected hyperparameters to optimize recognition performance in terms of \gls{WER} which does not translate to a good diarization performance.
In particular, the morphological closing used for smoothing in the \gls{VAD} overestimates speech activity for good \gls{WER} results \cite{Boeddeker2023_TSSEP}, hence our \gls{DER} cannot compete with the \gls{DER} parameters from \cite{Boeddeker2023_TSSEP}.
Nevertheless, we obtained a solid \gls{DER} of \SI{9.4}{\percent} on Libri-CSS.

Our proposed pipeline outperforms prior works in terms of cpWER. 
While the Whisper \gls{ASR} is necessary for the sub-segmentation, a second \gls{ASR} pass can be applied to the sub-segments after the full pipeline.
We used the ESPnet \gls{ASR} for the second ASR pass since it performs better on LibriSpeech data.
With that second ASR pass, we obtained a further improvement of \SI{1}{\percent}, which outperforms the best single-channel \gls{cpWER} on Libri-CSS by a \SI{20}{\percent} relative cpWER improvement.

\begin{table}
    \centering
    \caption{cpWER on Libri-CSS using TF-GridNet, energy-based VAD, Whisper and different sub-segmentations.}
\sisetup{round-mode=places,round-precision=1}
\renewcommand*{\arraystretch}{1.1}
    \begin{tabular}{lSS}
        \toprule
        Sub-segmentation & \multicolumn{2}{c}{Clustering} \\
        \cmidrule{2-3}
         & {k-Means} & {Oracle} \\
         \midrule
        -- & 14.83 & \color{gray} 13.91 \\
        uniform 4s & 9.83 & \color{gray} 7.54 \\
        uniform 2s & 12.35 & \color{gray} 6.79 \\
        \midrule
        sentence & 7.859717087434696 & \color{gray} 7.353119208813313 \\
        word & 7.6 & \color{gray} 7.1 \\
        sentence + word & 7.2 & \color{gray} 6.6 \\
         \bottomrule
    \end{tabular}
    \label{tab:main}
\end{table}

\begin{table}[]
\centering
    \caption{Single channel Libri-CSS literature comparison
}
    \sisetup{round-mode=places,round-precision=1}
    \renewcommand*{\arraystretch}{1.2}
    \label{tab:der_literature}
    \begin{tabular}{lSS}
        \toprule
        System & {DER} & {cpWER} \\
        \midrule
         SC \cite{21_raj_scovl} & 11.28 & {--} \\
         RPN \cite{huang2022joint} & 9.5 & {--} \\
SC + TS-VAD + ESPnet \cite{Boeddeker2023_TSSEP} & 5.65 & 9.26 \\
         SC + TS-SEP + ESPnet \cite{Boeddeker2023_TSSEP} & 15.68 & 7.81 \\
         Transcribe-to-Diarize \cite{Kanda2022_TranscribeToDiarize} & 7.9 & 11.6 \\
         \midrule
         Proposed & 9.39 & 7.21 \\
         Proposed + ESPnet & 9.39 & 6.17 \\
         \bottomrule
    \end{tabular}
\end{table}

\section{Conclusions}
We presented a modular pipeline for transcribing a meeting that consists of a source separation, a recognition, and a diarization module that achieves state-of-the-art \gls{cpWER} performance on Libri-CSS. 
Using TF-GridNet as a separator in the CSS module delivers excellent separation results. 
Since a VAD is unable to detect fast speaker turns, we employed sentence boundary predictions by the ASR for diarization.
But even the ASR engine may miss sentence boundaries.
The proposed sub-segmentation based on word boundary information to fix this issue further improved the performance, and combining sentence and word information delivered a new state-of-the-art single-channel cpWER of \SI{7.2}{\percent} and \SI{6.2}{\percent} with Whisper and ESPnet, respectively.
Thus, to answer the question posed in the introduction, a modular pipeline with source separation followed by ASR and followed by diarization is highly competitive, even with single-channel input.

\section{Acknowledgements}
Computational Resources were provided by BMBF/NHR/PC2.
Christoph  Boeddeker was funded by DFG, project no.\ 448568305.

\bibliographystyle{IEEEbib}
\bibliography{refs}

\end{document}